\documentclass[conference]{IEEEtran}

\usepackage{graphicx}
\usepackage{amsmath,amssymb}
\usepackage{booktabs}
\usepackage{cite}

\title{Signal-Informed Temporal Routing for Vinyl Defect Regime Detection\\[7pt]
\small\begin{tabular}{@{}c@{}}
Columbia University Nonlinear Control Lab\\
Technical Report CUNLC-20260919-01
\end{tabular}}

\author{
\IEEEauthorblockN{
Yi-Hung Kan\textsuperscript{2}
and
Homayoon Beigi\textsuperscript{1,2,3,4}
}
\IEEEauthorblockA{
\textsuperscript{1}Nonlinear Control Laboratory,
\textsuperscript{2}Dept. of Electrical Engineering,
\textsuperscript{3}Dept. of Mechanical Engineering\\
Columbia University, New York, NY, USA\\
\textsuperscript{4}Recognition Technologies, Inc., South Salem, NY, USA\\
Email: yk3164@columbia.edu and hb87@columbia.edu
}
}

\begin{document}

\maketitle

\begin{abstract}
Vinyl restoration systems must distinguish isolated clicks, short bursts,
dense crackle, and overlapping damage before selecting a repair operation.
We present a lightweight two-stage detector in which signal-informed sparse,
burst, and dense experts produce complementary defect evidence, and a
temporal backend converts that evidence into stable repair regimes. The
backend factors the five-way decision hierarchically, applies a
validation-only mixed regime gate, and decodes with validation-selected
transition penalties that outperform a maximum-likelihood transition matrix
on the same emissions. On a source-separated synthetic benchmark
of 597 non-overlapping 15\,s excerpts drawn from 21 recordings, the held-out
system reaches 0.730 pooled five-class macro F1 (95\% CI 0.694--0.761),
an improvement of 0.043 over a flat frame-level router
(paired bootstrap $p=0.001$). Clean and dense frames are highly reliable,
while sparse, burst, and mixed frames remain far more ambiguous.
\end{abstract}


\section{Introduction}

Archival audio restoration is difficult because degradation is not a single
noise process. A digitized record can contain isolated clicks, wide pops, short scratches,
dense crackle, low-frequency disturbances, and overlapping damage. These artifacts differ in duration, density, spectrum, and repair assumption.
A narrow click may be repaired by local interpolation, whereas a scratch
requires region inpainting and dense crackle a different denoising strategy. A restoration system therefore needs to detect damage and, at the same time,
route each time region to the appropriate repair mode.

Classical declicking assumes a localized impulsive corruption model,
\begin{equation}
    x[n]=s[n]+c[n],
\end{equation}
where $s[n]$ is clean audio and $c[n]$ a sparse impulsive component
\cite{godsill1998}. This motivates detect-and-replace pipelines but becomes
unreliable when micro impulses occur densely, defects overlap, or musical
transients resemble damage. A single declicking score cannot decide whether a
region should be interpolated, inpainted, denoised, or left unchanged.

This paper formulates vinyl defect analysis as a repair-oriented regime
routing problem. The front end produces three evidence streams, namely candidate-level sparse
evidence for isolated clicks and pops, frame-level burst evidence for wide
pops and scratches, and frame-level dense evidence for crackle. A learned hierarchical backend then combines these streams with confidence,
interaction, and local temporal features, and a validation-selected Viterbi
decoder produces a stable five-class sequence over clean, sparse, burst,
dense, and mixed regimes.

Our contributions are fourfold. First, we design a two-stage defect-regime
detector whose front end pairs signal-informed candidate and frame features
with three task-specific experts. Second, we show that factoring the decision
hierarchically and decoding it with validation-selected transition penalties
improves macro F1 over both a flat router and a maximum-likelihood transition
matrix, under both the five-class taxonomy and the coarser four-class
taxonomy that a repair stage consumes. Third, we establish by permutation
analysis that the expert evidence streams, rather than the raw frame
descriptors available to the same router, carry the routing decision.
Fourth, we quantify the uncertainty of every reported gain with a
case-level paired bootstrap and report per-recording variation, and we
release a reference implementation that reproduces every number.

\section{Related Work}

Classical restoration models clicks, pops, and scratches as localized
impulsive disturbances. Statistical and autoregressive methods use prediction
error or local signal models to find corrupted samples and then interpolate
or suppress them \cite{godsill1998,esquef2002}, and bidirectional processing
improves pulse localization by combining forward and backward alarms
\cite{niedzwiecki2013}. These motivate our event-centered sparse branch, but
their output is a repair mask rather than complementary regime evidence.

Supervised archival detection moves the decision earlier. Brandt et al.
classify whether each 1\,s frame contains impulsive disturbance, using
prewhitening, so the resulting probabilities can gate a restoration
algorithm and leave clean frames untouched \cite{brandt2017}. Prior work
separates sporadic clicks, crackle as a dense floor of small clicks, and
low-frequency thumps, noting that dense bursts are perceived as crackle
\cite{brandt2018thesis,moliner2022diffusion}. These temporal structures
motivate our event-level sparse branch and frame-level burst and dense
branches, extending binary gating to repair-oriented regimes.
Perceptual-model detectors have likewise localized clicks in real analog
excerpts \cite{rund2016}.

Our feature choices also relate to transient analysis. Abrupt energy,
spectral, and local statistical changes mark musical onsets as well as noise
events \cite{bello2005}, the central difficulty here, since high curvature,
prediction error, or high-frequency energy is informative without proving a
defect. Our front end therefore pairs LPC residual and curvature cues with
shape, spectral, cepstral, density, and context statistics, and keeps
candidate-level localization for the backend to interpret.

The repair target also determines what counts as useful detection. Audio
inpainting reconstructs corrupted samples from surrounding structure
\cite{adler2012}, whereas crackle and overlapping damage need region-level
denoising. Antiquing supplies a physical basis for synthesizing historical
degradations \cite{valimaki2008,moliner2022diffusion} and neural systems
learn from them directly \cite{moliner2022unet}, but both target restoration
quality rather than prerestoration routing. Our contribution is the
intermediate representation, in which physically motivated experts produce
separate evidence for later regime selection without forcing every artifact
type through a single detector.

\section{Methodology}
\label{sec:methodology}

The detector has two stages. A signal-informed front end estimates sparse,
burst, and dense evidence, and a backend router then predicts
\begin{equation}
    y_t \in \{\mathrm{clean},\mathrm{sparse},\mathrm{burst},
    \mathrm{dense},\mathrm{mixed}\}
\end{equation}
for each analysis frame $t$.

The two stages use different supervision units. Sparse damage is localized
around event centers, whereas bursts and crackle occupy short or extended
regions. Keeping these units separate preserves physical information, and the
backend is the first stage that must emit one exclusive label per frame.

\subsection{Benchmark Construction and Frame Targets}

Each excerpt is corrupted by a seed-controlled recipe and stored with its
clean and corrupted waveforms, sample-level masks, event table, and frame
labels, giving both an inference input and an audit trail. Excerpt boundaries
are fixed before corruption, and all feature and label arrays reset there.

The sparse branch is trained at candidate level, where a candidate is positive
when its local maximum lies within 100 samples of a ground-truth event center.
Burst and dense supervision are frame level and follow the supported-artifact
logic for wide pops and short scratches and the dense mask-support criterion.
For router training, candidate-level sparse evidence is projected onto frames.
A frame is clean when no supported defect is active, mixed when dense and
impulsive support coincide, and otherwise sparse, burst, or dense.

\subsection{Signal-Informed Expert Front End}

The waveform is converted to mono, median-centered, and normalized by a
median absolute deviation estimate. An order-16 LPC residual and a
second-difference curvature signal are normalized the same way, and their
positive components form the core defect score
\begin{equation}
    s_{\mathrm{core}}[n]=
    \sqrt{\max(s_{\mathrm{LPC}}[n],0)\max(s_{\Delta^2}[n],0)}.
\end{equation}
Local maxima define sparse candidates, described by residual and curvature
magnitude, prominence, width, spacing, and density. Frame features use
1024-sample frames with a 512-sample hop and cover spectral flatness,
high-frequency ratio, spectral spread, cepstral energy, crest factor,
core-score statistics, peak count, and peak density.

Three binary MLP experts estimate complementary evidence, each with two ReLU
hidden layers of 32 and 16 units, a sigmoid output, and standardization
fixed from the training pool. The sparse expert operates on candidates and
the burst and dense experts on frames. Their objectives are trained
independently, so a strong response in one branch does not suppress another.
Sparse candidate probabilities are aggregated per frame by the maximum,
giving
\begin{equation}
    \mathbf{p}_t=[P_s(t),P_b(t),P_d(t)]^T.
\end{equation}
The outputs are evidence streams rather than mutually exclusive class
probabilities, and mixed frames arise when dense support co-occurs with
impulsive activity.

\subsection{Hierarchical Temporal Router}

For each frame the router receives a 101-dimensional vector
\begin{equation}
    \mathbf{z}_t = [\mathbf{p}_t^T,\;\mathbf{c}_t^T,\;
    \mathbf{i}_t^T,\;\mathbf{u}_t^T,\;\mathbf{g}_t^T]^T,
\end{equation}
where $\mathbf{c}_t$ holds the two largest expert probabilities, their
margin, and the normalized posterior entropy, $\mathbf{i}_t$ holds pairwise
expert differences and an agreement term, $\mathbf{u}_t$ holds local temporal
statistics, and $\mathbf{g}_t$ holds raw frame descriptors together with
candidate-density counts and their one-frame lag and lead. The temporal
statistics are rolling means, standard deviations, extrema, first and second
differences, and transition density of the expert trajectories over a
centered window of $\pm 4$ frames. The router is therefore non-causal by up
to 46\,ms, which suits an offline prerestoration pass. All windows are
computed within an excerpt, so context cannot cross unrelated source segments.

Rather than one five-way classifier, the backend factors the decision into
three softmax stages mirroring the physical taxonomy,
\begin{equation}
q_t(k)=
\begin{cases}
P(\neg\mathrm{def}), & k=0,\\
P(\mathrm{def})\,P(\tau_k\mid\mathrm{def})\,P(k\mid\tau_k), & k>0,
\end{cases}
\end{equation}
where stage one separates clean from defective frames, stage two assigns a
defect type $\tau\in\{\mathrm{impulsive},\mathrm{dense},\mathrm{mixed}\}$,
and stage three splits impulsive into sparse and burst, with
$P(k\mid\tau_k)=1$ for the dense and mixed types. The stages are ReLU MLPs with 28, 28, and 16 hidden units, each trained with
class-balanced cross entropy on a capped per-class subsample, so the two
majority regimes cannot dominate the macro-averaged objective. This lets the
rare sparse/burst distinction be learned on impulsive frames alone rather
than against 413{,}385 clean frames.

\subsection{Mixed Calibration and Sequence Decoding}

Mixed damage is treated as an overlap state rather than an independent
morphology. The backend combines dense evidence with impulsive evidence
\begin{equation}
    e_{\mathrm{imp}}(t)=\max(P_s(t),P_b(t)),
\end{equation}
and suppresses the mixed posterior by $0.02$ on any frame where
$P_d(t)<0.75$ or $e_{\mathrm{imp}}(t)<0.65$. A second pass repeats this
wherever the mixed posterior fails to exceed the dense posterior by $-0.12$,
removing mixed hypotheses that survive only because dense evidence is strong.
Both thresholds are fixed on validation data. Given the
resulting emissions $\widetilde q_t(k)$, the final sequence minimizes
\begin{equation}
 J(\mathbf{y})=\sum_t -\log(\widetilde q_t(y_t)+\epsilon)
 +\sum_{t>1}C(y_{t-1},y_t)+\sum_tB(y_t),
\end{equation}
where $C$ penalizes changes and selected unstable transitions, and $B$ is a
per-frame class bias that discourages the mixed and isolated impulsive states.

The sequence follows from the standard Viterbi recursion. If $D_t(k)$ is the
minimum cost of a partial sequence ending in state $k$, then
\begin{equation}
 D_t(k)=-\log(\widetilde q_t(k)+\epsilon)+B(k)
 +\min_j\{D_{t-1}(j)+C(j,k)\}.
\end{equation}
Backpointers recover the minimum-cost sequence independently per excerpt, so
the decoder suppresses isolated low-confidence flips while keeping a
transition when accumulated emission evidence supports it.

All router, calibration, and decoder parameters are selected on a validation
split of the training pool. The stay bonus is 1.50, switch penalty 0.25,
mixed prior penalty 0.30, sparse/burst prior penalty 0.25, and the
dense/mixed and clean/defect transition penalties both 0.10. Selection
follows a tune-then-refit protocol. The penalties and mixed gate are chosen
with a hierarchy fitted on the 418 fitting cases and scored on the 70
validation cases, then frozen, after which the hierarchy is refitted on all
488 development cases. The decoder is thus frozen before any held-out label
is read, and test labels serve only the final metrics and the confusion
matrix. The penalties therefore act on emissions from a model refitted on
more data than selected them, which we revisit in
Section~\ref{sec:conclusion}.

\section{Experiments and Results}
\label{sec:results}

\subsection{Benchmark and Protocol}

Because real transfers cannot supply sample-accurate labels, we use a
controlled synthetic benchmark built from clean music recordings, divided
into deterministic non-overlapping 15\,s excerpts at 44.1\,kHz. Each realization injects one of five procedural recipes, namely narrow
clicks, wide pops, short scratches, dense crackle, or mixed damage. The split is made at the level of the source recording. Eighteen development
recordings yield 488 realizations and three held-out recordings yield 109,
for 597 in total. The development pool is split by recording into 418 fitting
and 70 validation cases. The held-out recordings are piano, orchestral, and
vocal-jazz material sharing no source with the development pool. Their small
number is the main limit on generalization, so every headline figure carries
a case-level bootstrap interval.

All expert outputs are frozen for the backend comparison, and no held-out
label is used for fitting, calibration, threshold selection, transition
penalties, or decoder selection.
We report two taxonomies. The five-class view separates sparse from burst
and is the harder diagnostic target. The four-class view merges them into one
impulsive class and matches what a repair stage consumes, since isolated
clicks and short bursts are both handled by interpolation. Both are scored as
pooled macro F1 over the aggregate confusion matrix, with weighted F1 also
reported because clean and dense dominate the frames.
Uncertainty is estimated by resampling whole excerpts with replacement over
4000 draws, which respects the correlation between frames inside an excerpt.
System comparisons use the paired difference on the same resampled excerpts.
No validation candidate satisfied the auxiliary clean, dense, and mixed
floors at once. Once that set proved infeasible, the final rule was selected
by the predefined validation-only penalized objective. This decision was made before any held-out test labels were read.

\subsection{Frozen Expert Front End}

Table~\ref{tab:expert_results} summarizes standalone expert behavior. Wide
pops activate both the sparse and burst experts, short scratches mainly
burst, and dense crackle is isolated cleanly by the dense expert. Mixed
damage retains strong dense evidence while also activating both impulsive
streams.

\begin{table}[t]
\centering
\caption{Frozen expert results on held-out TEST. Sparse F1 is candidate/event
level. Burst and dense F1 are frame level. A zero entry means the recipe
contains no positives of that type.}
\label{tab:expert_results}
\scriptsize
\begin{tabular}{lccc}
\toprule
Defect recipe & Sparse F1 & Burst F1 & Dense F1 \\
\midrule
Narrow clicks & 0.605 & 0.000 & 0.000 \\
Wide pops & 0.927 & 0.899 & 0.000 \\
Short scratches & 0.000 & 0.634 & 0.000 \\
Dense crackle & 0.000 & 0.000 & 0.986 \\
Mixed damage & 0.577 & 0.378 & 0.975 \\
\bottomrule
\end{tabular}
\end{table}

Overall sparse candidate F1 is 0.714, burst frame F1 is 0.511, and dense
frame F1 is 0.961. A direct front-end evidence rule nonetheless reaches only
0.464 pooled macro F1 and produces 14{,}725 transitions, so the expert front
end is physically meaningful but insufficient as a complete repair router.

\subsection{Backend Results}

Table~\ref{tab:backend_results} reports the main held-out results. Direct
thresholding of expert evidence is far weaker than learned routing. A flat
five-class router reaches 0.687 macro F1, and the hierarchy, temporal
consistency, and Viterbi decoding raise this to 0.702, 0.713, and 0.730
(95\% CI 0.694--0.761). The gain of the full backend over the flat router is $+0.043$
(95\% CI $+0.016$ to $+0.070$, $p=0.001$). Decomposing it, the hierarchical
factorization alone contributes $+0.015$, which the paired test cannot
separate from zero (95\% CI $-0.008$ to $+0.039$, $p=0.20$), whereas
calibration and sequence decoding contribute a further $+0.028$
(95\% CI $+0.009$ to $+0.045$, $p=0.003$). The practical gain therefore comes
from temporal decoding rather than from the hierarchy on its own.

The same ordering holds under the four-class taxonomy, where the final system
reaches 0.805 (95\% CI 0.777--0.829) against 0.768 for the flat router, a
gain of $+0.037$ ($p<0.001$). That interval is the narrower of the two,
because merging the two smallest-support classes leaves the repair-relevant
claim better determined than the diagnostic one.

A natural alternative is to estimate the transition matrix directly.
Decoding the same emissions with maximum-likelihood transitions and priors
from the training labels gives 0.712 macro F1, below the 0.730 from
validation-selected penalties (paired difference $+0.018$, 95\% CI $+0.002$
to $+0.035$, $p=0.025$). The
maximum-likelihood decoder is the more conservative of the two, emitting
1{,}020 transitions against 1{,}476 and reaching a slightly better mixed F1,
but it oversmooths sparse to 0.447. Empirical transition counts are dominated
by the two majority regimes, so a small penalty set tuned against a
macro-averaged objective transfers better than the likelihood fit.

\begin{table*}[t]
\centering
\caption{Held-out TEST performance under both taxonomies, cumulative backend
stages, all with frozen expert evidence and validation-only selection. Final
row 95\% bootstrap CI: 0.694--0.761 five-class, 0.777--0.829 four-class.}
\label{tab:backend_results}
\setlength{\tabcolsep}{8pt}
\begin{tabular}{lcccccc}
\toprule
 & \multicolumn{4}{c}{Five-class} & 4-cl. & \\
\cmidrule(lr){2-5}
 & Macro & Wtd. & Micro & Mixed & Macro & \\
System & F1 & F1 & F1 & F1 & F1 & Trans. \\
\midrule
Front-end evidence rule & 0.464 & 0.926 & 0.895 & 0.051 & 0.597 & 14725 \\
Flat five-class router & 0.687 & 0.980 & 0.978 & 0.413 & 0.768 & 2439 \\
+ hierarchy and mixed gate & 0.702 & 0.982 & 0.981 & 0.335 & 0.775 & 2419 \\
+ temporal consistency & 0.713 & 0.983 & 0.982 & 0.367 & 0.785 & 2119 \\
\midrule
Viterbi, ML transitions & 0.712 & 0.984 & 0.984 & 0.438 & 0.796 & 1020 \\
Viterbi, selected penalties & \textbf{0.730} & \textbf{0.983} & \textbf{0.983} & 0.422 & \textbf{0.805} & 1476 \\
\bottomrule
\end{tabular}
\end{table*}

A remaining question is whether the experts carry the routing information or
whether the raw descriptors in $\mathbf{g}_t$ suffice alone. Permuting each
feature group across frames and re-decoding over five repeats, the 50
expert-derived coordinates cost 0.519 macro F1 when destroyed, collapsing the
system to 0.211, against 0.095 for the 39 raw descriptors and 0.074 for the
12 candidate counts. The expert streams are what the backend routes on, and
the descriptors act as secondary context rather than a substitute, which is
why the front end is worth training separately.

\subsection{Per-Class Performance and Failure Analysis}

Table~\ref{tab:per_class} gives the final per-class metrics. Clean and dense
reach 0.991 and 0.980 F1, burst 0.691 and sparse 0.565, and mixed remains
the limiting class at 0.422.

\begin{table}[t]
\centering
\caption{Final Viterbi per-class metrics on held-out TEST.}
\label{tab:per_class}
\scriptsize
\begin{tabular}{lrrrr}
\toprule
Regime & Precision & Recall & F1 & Support \\
\midrule
Clean & 0.995 & 0.987 & 0.991 & 92788 \\
Sparse & 0.599 & 0.534 & 0.565 & 442 \\
Burst & 0.627 & 0.770 & 0.691 & 478 \\
Dense & 0.973 & 0.987 & 0.980 & 46465 \\
Mixed & 0.413 & 0.432 & 0.422 & 437 \\
\bottomrule
\end{tabular}
\end{table}

\begin{figure}
    \centering
    \includegraphics[width=0.84\linewidth]{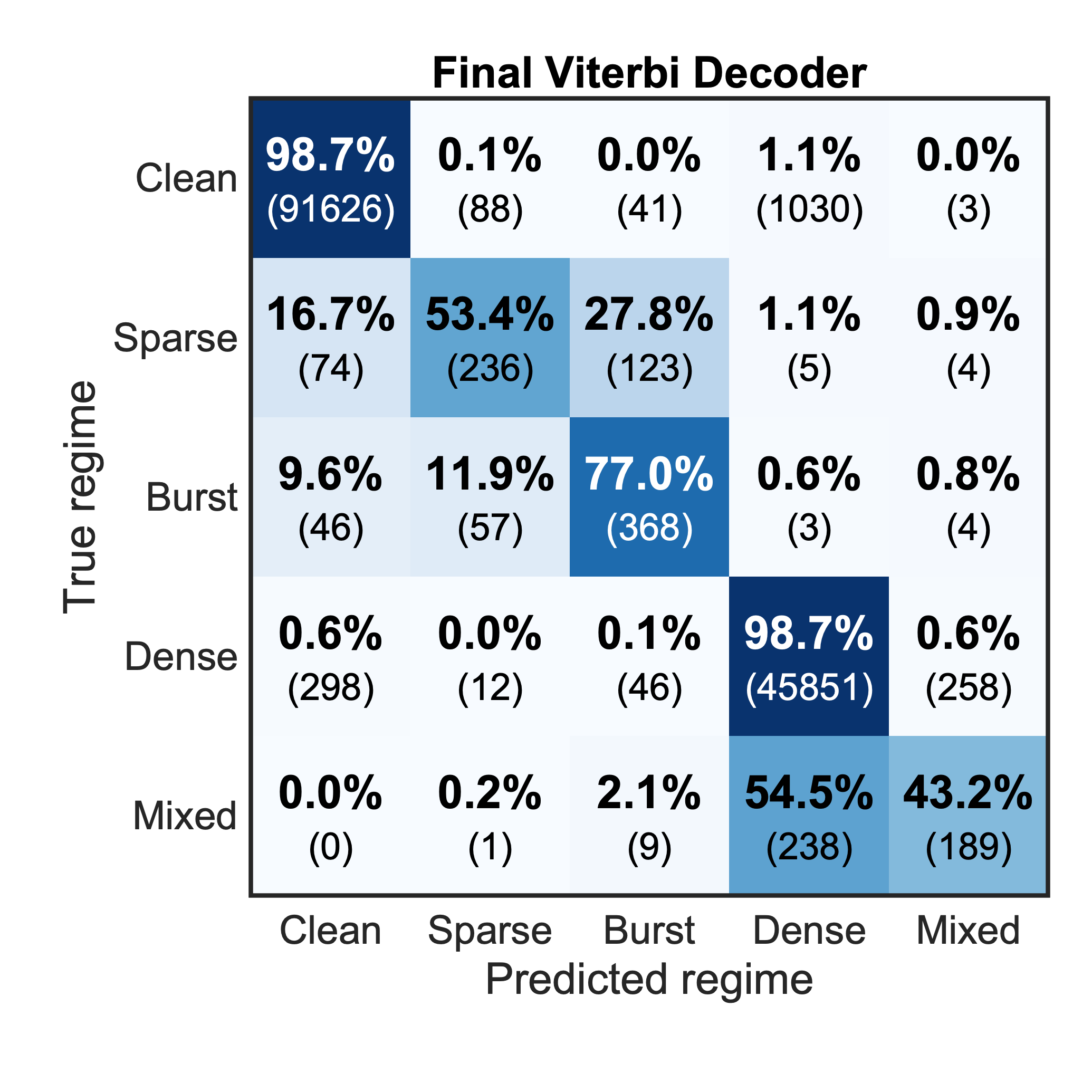}
    \caption{Row-normalized held-out TEST confusion matrix for the final
    Viterbi-decoded router. Clean and dense regimes are reliable. Mixed errors
    are dominated by confusion with dense, while sparse and burst remain
    partially ambiguous.}
    \label{fig:confusion}
\end{figure}

Fig.~\ref{fig:confusion} shows the remaining errors are structured. Sparse
frames are taken for burst in 27.8\% of cases and burst for sparse in
11.9\%. Of mixed frames, 54.5\% are predicted dense and 43.2\% correctly,
whereas only 0.6\% of dense frames are predicted mixed. Mixed therefore
carries dense evidence plus impulsive activity, unlike both a clean fifth
morphology and an ordinary dense frame. It is also the least certain class,
with 437 frames and a 95\% interval of 0.352--0.494.
Much of this comes from the label space rather than the features. Merging
sparse and burst lifts impulsive F1 to 0.825, above either constituent
class, so the evidence separating impulsive from dense activity is far
stronger than that separating the two impulsive morphologies.

Temporal decoding cuts transitions from 14{,}725 under direct thresholding
to 1{,}476, and from 2{,}119 relative to temporal consistency alone while
macro F1 rises from 0.713 to 0.730, so the decoder improves class
interpretation as well as sequence stability.

\subsection{Generalization Across Recordings}

Because the held-out pool is small, Table~\ref{tab:per_track} breaks the final
result down by source recording. Macro F1 ranges from 0.656 to 0.765 under
the five-class taxonomy and from 0.740 to 0.819 under the four-class
taxonomy. The weakest case is the vocal-jazz excerpt set, which is also the
smallest at twelve realizations, and the strongest is the orchestral piano
concerto. The spread is wide enough that a single additional recording could
move the pooled figure by more than the gains reported above, which is why we
treat the bootstrap intervals rather than the point estimates as the claim.

\begin{table}[t]
\centering
\caption{Final system by held-out source recording.}
\label{tab:per_track}
\footnotesize
\begin{tabular}{lccc}
\toprule
Source recording & Cases & 5-class & 4-class \\
\midrule
Chopin, Ballade No.\ 4 & 50 & 0.712 & 0.803 \\
Rachmaninoff, Piano Concerto No.\ 2 & 47 & 0.765 & 0.819 \\
Fitzgerald and Armstrong, vocal jazz & 12 & 0.656 & 0.740 \\
\midrule
Pooled & 109 & 0.730 & 0.805 \\
\bottomrule
\end{tabular}
\end{table}

\subsection{Reproducibility}

The decoder, the calibration stages, and the evaluation are deterministic
given the frozen models, so every number in this paper is reproducible from
the released artifacts. A reference implementation independent of the
original MATLAB code reproduces all reported metrics to within
$5\times10^{-16}$, and it also regenerates the bootstrap intervals, the
maximum-likelihood transition baseline, and the permutation analysis.

\section{Conclusion}
\label{sec:conclusion}

We presented a lightweight signal-informed system for prerestoration vinyl
defect regime detection. Frozen sparse, burst, and dense expert evidence
feeds a hierarchical router, a mixed-regime gate, and a validation-selected
Viterbi decoder, reaching 0.730 five-class and 0.805 four-class macro F1,
ahead of a flat router by 0.043 and 0.037. A paired bootstrap attributes most
of that gain to calibration and sequence decoding rather than the hierarchy,
and the selected penalties also beat a maximum-likelihood transition matrix.
Temporal structure rather than classifier capacity makes the evidence
usable.

Four limits bound these claims. The taxonomy treats overlap as exclusive,
tune-then-refit applies penalties to emissions from a refitted hierarchy,
a held-out pool of three recordings is a narrow
basis for generalization, and
procedural degradation is not real vinyl wear. The first three need more data
and a stricter protocol. The fourth is harder, because real transfers cannot
supply the frame-level ground truth this evaluation depends on, so validation
there needs region-level expert annotation and a downstream comparison in
which routing selects the repair operator and is judged against a
single-operator baseline. Transfer to an independent degradation model is a
cheaper intermediate check.

\end{document}